\documentclass[11pt]{article}
\usepackage{amssymb}
\usepackage{amsmath}
\usepackage{amscd}
\usepackage{latexsym}

\catcode`@=11
\renewcommand{\section}{\@startsection{section}{1}{0pt}{\medskipamount}
{\medskipamount}{\large\bf}}
\numberwithin{equation}{section}
\catcode`@=12
\def\a{\alpha}

\def\g{\gamma}

\def\de{\delta}

\def\la{\lambda}

\def\o{\omega}

\newcommand{\Ncal}{{\cal N}}
\newcommand{\Lcal}{{\cal L}}

\newcommand{\with}{{\quad{\rm with}\quad}}
\newcommand{\for}{{\qquad{\rm for}\qquad}}
\renewcommand{\and}{{\quad{\rm and}\quad}}
\newcommand{\und}{{\qquad{\rm and}\qquad}}

\def\im{\mbox{i}}
\def\pa{\mbox{$\partial$}}
\def\diff{\mbox{d}}

\def\sfrac#1#2{{\textstyle\frac{#1}{#2}}}
\def\>{\rangle}
\def\<{\langle}
\def\+{\dagger}
\def\={\ =\ }

\def\Qr{{\buildrel{Q}\over{\longrightarrow}}}
\def\Ql{{\buildrel{Q}\over{\longleftarrow}}}

\begin{document}
\begin{titlepage}
\setcounter{page}{0}
\begin{flushright}
ITP--UH--06/13\\
CBPF-NF-001/13
\end{flushright}

\vskip 1.5cm

\begin{center}

{\Large\bf 
Target duality in ${\cal N}{=}\,8$ superconformal mechanics \\[12pt]
and the coupling of dual pairs
}

\vspace{12mm}
{\large Marcelo Gonzales${}^\+$, Sadi Khodaee${}^*$, 
Olaf Lechtenfeld${}^{\times\circ}$ and Francesco Toppan${}^*$}\\[8mm]

\noindent ${}^\+${\em Carrera de F\'isica\\
Universidad Aut\'onoma Tom\'as Fr\'ias\\
Av.\ Del Maestro s/n, Casilla 36, Potos\'i, Bolivia} \\
{Email: marcbino@gmail.com}\\[6mm]

\noindent ${}^*${\em TEO, CBPF\\
Rua Dr. Xavier Sigaud 150 (Urca)\\
Rio de Janeiro (RJ), cep 22290-180, Brazil} \\
{Emails: khodaee@cbpf.br, toppan@cbpf.br}\\[6mm]

\noindent ${}^\times${\em
Institut f\"ur Theoretische Physik and Riemann Center for Geometry and Physics\\
Leibniz Universit\"at Hannover \\
Appelstra\ss{}e 2, 30167 Hannover, Germany }\\
{Email: lechtenf@itp.uni-hannover.de}\\[6mm]

\noindent ${}^{\circ}${\em
Centre for Quantum Engineering and Space-Time Research\\
Leibniz Universit\"at Hannover \\
Welfengarten 1, 30167 Hannover, Germany }
\vspace{12mm}

\begin{abstract}
\noindent
We couple dual pairs of ${\cal N}{=}\,8$ superconformal mechanics with
conical targets of dimension $d$ and~$8{-}d$. The superconformal coupling
generates an oscillator-type potential on each of the two target factors,
with a frequency depending on the respective dual coordinates.
In the case of the inhomogeneous (3,8,5) model, which entails a monopole
background, it is necessary to add an extra supermultiplet of constants for 
half of the supersymmetry. The ${\cal N}{=}\,4$ analog, joining an
inhomogeneous (1,4,3) with a (3,4,1) multiplet, is also analyzed in detail.
\end{abstract}
\end{center}

\end{titlepage}

\section{Introduction and summary}

\noindent
${\cal N}$-extended superconformal mechanics (for a review, see~\cite{rev})
is defined on off-shell supermultiplets containing propagating bosons, fermions and auxiliary fields and, 
following the conventions of~\cite{pt}, being denoted by $(d, {\cal N}, {\cal N}{-}d)$.
Their associated invariant actions define one-dimensional sigma models 
with a $d$-dimensional conical target manifold. 
The case of ${\cal N}{=}8$ has been studied less extensively than those with ${\cal N}{=}2$ or ${\cal N}{=}4$.
However, in the literature one finds invariant actions for the supermultiplets
(1,8,7)~\cite{krt,di}, (3,8,5)~\cite{bikl1,bikl2} and (5,8,3)~\cite{bikl1,de}. 
The (2,8,6)~model is free.

In this paper, we make use of a $d\leftrightarrow {\cal N}{-}d$ duality observed in~\cite{khto} 
to couple for the first time two dually related superconformal mechanics. 
Depending on the target dimension $d$, for ${\cal N}{=}8$ the coupled systems are invariant under 
one of the four one-dimensional finite superconformal algebras $A(3,1)$, $D(4,1)$, $D(2,2)$ or~$F(4)$. 
Their target manifold is a product of two asymptotically flat cones of dimension~$d$ and $8{-}d$ over
the spheres $S^{d-1}$ and $S^{7-d}$, respectively.

The possibility of consistently coupling dually related supermultiplets was first observed, 
for homogeneous supersymmetry transformations, in~\cite{top}. 
This produces ${\cal N}{=}8$ superconformal systems with targets of dimension $d=1{+}7$, $2{+}6$ 
or $3{+}5$ (the 4-dimensional system is degenerate, and the dual of the 8-dimensional system is empty).
However, for the particular cases of $({\cal N}{=}4,d{=}1)$ and $({\cal N}{=}8,d{=}3)$,
an inhomogeneous deformation of the supersymmetry is admissible 
(see, e.g., \cite{kuto} and~\cite{khto}, respectively).
The presence of an inhomogeneity parameter is responsible for the appearance of a Calogero potential 
in the ${\cal N}{=}4$, $A(1,1)$-invariant, (1,4,3)~model and of a Dirac monopole in the 
${\cal N}{=}8$, $D(2,1)$-invariant,  (3,8,5)~model, as will be reviewed below. 
In these instances, a consistent superconformal coupling of the inhomogeneous supermultiplet
with its (homogeneous) dual is non-trivial, as will be shown here.
It requires the introduction of an extra supermultiplet of constants for half of the supersymmetries
and leads to new superconformal interactions in the presence of a Calogero potential or a monopole.
In particular, in all cases (homogeneous or not), an oscillator potential on each of the two cones
is generated, with a frequency depending on the mutually dual coordinate.

The description of the models is given in a Lagrangian framework. 
By setting all fermionic fields to zero and eliminating the auxiliary fields, we are led to the dynamics 
of two interacting bosonic sigma models whose parameters are fixed by superconformal invariance.
Passing to conical radial variables then reveals the geometry and the physical content of the coupled model.
In this fashion, our results provide an extension of the class of known superconformal models.

Some interesting questions are left for future investigations. 
In particular, it seems quite plausible that the bosonic sector of the dually coupled models, 
whose parameters are fixed by superconformal invariance, turn out to be 
integrable, as a remnant of the off-shell invariant transformations. 

The paper is structured as follows.
After reviewing general features of $(d,8,8{-}d)$ supermultiplets in Section~2,
we present in Section~3 the superconformal pairing of dually related multiplets 
and work out the coupled Lagrangian in the case of homogeneous supersymmetry, 
ending up with the general bosonic potential on the cone product in the presence of Fayet-Iliopoulos terms.
Sections 4 and~5 deal with the inhomogeneous (3,8,5) supermultiplet, its Dirac monopole background 
and the corresponding gauge transformations. In Section~6, the dual (5,8,3) supermultiplet is displayed,
before Section~7 couples it to the inhomogeneous (3,8,5)~model. Here one finds the central results of
the paper. In Section~8 we reduce the coupled system back to the (5,8,3) supermultiplet.
Complete actions and the ${\cal N}{=}4$ coupling of the inhomogeneous (1,4,3) supermultiplet 
with its dual (3,4,1) partner are presented in detail in three Appendices.

\bigskip

\section{Generalities for ($d$,$8$,$8{-}d$) supermultiplets}

\noindent 
$\Ncal{=}8$ superconformal mechanical systems realize the one-dimensional global 
supersymmetry algebra
\begin{equation}
\bigl\{ Q_i, Q_j\bigr\} \= 2\de_{ij} H 
\quad\with\quad i,j=1,\ldots,8 \und H \= \pa_t\ ,
\end{equation}
where $t$ parametrizes the particle worldline.
The corresponding supermultiplets are denoted by ($d$,$8$,$8{-}d$),
indicating $d$ propagating bosonic, $8$ propagating fermionic and $8{-}d$
auxiliary bosonic coordinate functions for the superparticle, which thus moves
on some $d$-dimensional target space parametrized by 
$x=\{x^a\,|\,a=1,\ldots,d\}$. 

In the construction of ${\cal N}{=}8$ superconformal actions we can make manifest
at most four of the eight supersymmetries. 
Picking by convention $Q_1$, $Q_2$, $Q_3$ and $Q_8$,
an $\Ncal{=}4$ invariant action reads
\begin{equation}
S_d \= \int\!\diff t\ \Lcal_d \= 
\int\!\diff t\ Q_8 Q_1 Q_2 Q_3 F(x)\ ,
\end{equation}
where $F(x)$ is a yet unconstrained function of all coordinates.
The restriction to the manifest ${\cal N}{=}4$ superalgebra splits the ${\cal N}{=}8$
supermultiplet,
\begin{equation}
(d,8,8{-}d) \ \longrightarrow\ (d_1,4,4{-}d_1)\ \oplus\ (d_2,4,4{-}d_2)
\quad\with d_1,d_2\le 4 \and d_1+d_2=d
\end{equation}
and opposite chiralities.\footnote{
The construction fails if $d_1=0$ or $d_2=0$. There exists, however, a different
method which works in all cases~\cite{gkt}.}
It turns out that the action depends only on two combinations of second derivatives
of~$F$, namely~\footnote{
Except for $d{=}3$, where an inhomogeneous deformation yields a background 
gauge potential, see below.}
\begin{equation}
\Phi_1\=-\Delta_{d_1} F\ \equiv\ -F_{1\,1}-\ldots-F_{d_1d_1} \und 
\Phi_2\=\Delta_{d_2} F\ \equiv\ F_{d_1+1\,d_1+1}+\ldots+F_{d\,d}\ ,
\end{equation}
where we grouped the coordinates according to the decomposition above.

To enhance to $\Ncal{=}8$ invariance, we must impose
\begin{equation}
Q_\ell \Lcal_d = \pa_t W_\ell \for \ell=4,5,6,7\ .
\end{equation}
This produces a harmonicity condition on~$F$,
\begin{equation}
\Delta_d F \ \equiv\ \delta^{ab}F_{ab} \= 0\ .
\end{equation}
As a consequence, we have
\begin{equation}
\Phi_1\=\Phi_2\ =:\ \Phi \quad\with\quad \Delta_d\Phi=0\ .
\end{equation}
Clearly, for $d\le5$, we may take $d_2=1$ so that $\Phi=F_{dd}$,
singling out the last coordinate.
Hence, taking $F$ to be harmonic, we obtain an ${\cal N}{=}8$ sigma model,
with a conformally flat target space for the propagating bosonic coordinates,
\begin{equation}
\diff s^2 \= \Phi(x)\,\delta_{ab}\diff{x}^a\diff{x}^b\ .
\end{equation}

The remaining generators of the conformal $sl(2)$ algebra
are realized as
\begin{equation}
K \= -t^2\pa_t-2t\,\la_\varphi \und D \= -t\,\pa_t-\la_\varphi
\end{equation}
on functions $\varphi$ with engineering dimension $[\varphi]=\la_\varphi$.
They give rise to 8 superconformal generators~\cite{kuto,khto}
\begin{equation}
\widetilde{Q}_i \= [K,Q_i]\ .
\end{equation}
Superconformal symmetry is imposed by also demanding that~\footnote{
The `$D$ condition' actually follows from the `$K$ condition'.}
\begin{equation}
D\,\Lcal_d \= \pa_t M_D \und
K\,\Lcal_d \= \pa_t M_K \ ,
\end{equation}
which yields two conditions on~$\Phi$, namely
\begin{equation}
[\Phi]=-1-2\la_x \und \Phi=\Phi(r)
\with r^2 \= \delta_{ab} x^a x^b\ .
\end{equation}
The closure of the D-module representation for the $\Ncal{=}8$ superconformal algebra
determines a critical value for the engineering dimension of~$x$,
\begin{equation}
\la_x \= \sfrac1{d-4} \qquad\Rightarrow\qquad 
[\Phi]\=-1-\sfrac2{d-4}\=\sfrac{d-2}{4-d}\=(2{-}d)\la_x\ .
\end{equation}
As a consequence, the conformal factor is indeed fixed to the proper harmonic expression,
\begin{equation}
\Phi(r) \= r^{2-d}\ .
\end{equation}

Introducing the spherical line element $\diff\Omega_{d-1}$ on $S^{d-1}$ and
changing the radial coordinate via
\begin{equation}
\rho \= \sfrac{2}{|4{-}d|} r^{(4-d)/2}\ ,
\end{equation}
the metric reads
\begin{equation}
\diff s^2 \= r^{2-d} \bigl( \diff r^2 + r^2\diff\Omega_{d-1}^2\bigr)
\= \diff\rho^2 + \sfrac14(4{-}d)^2\rho^2\diff\Omega_{d-1}^2\ .
\end{equation}
It reveals the target space to be a specific cone over $S^{d-1}$, 
asymptotically flat with a linear relative deficit of $|4{-}d|/2$.
Its scalar curvature comes out as
\begin{equation}
R \= \sfrac14(d{-}1)(d{-}2)^2(d{-}6)r^{d-4} 
\= (d{-}1)(d{-}2)^2(d{-}6)(d{-}4)^{-2}\rho^{-2}\ ,
\end{equation}
which is negative for $d=3,5$ and positive for $d=7,8$.
At $d=2,6$ we encounter flat space.

In any dimension~$d$ up to 8, the manifest $\Ncal{=}4$ superconformal algebra
must be a particular member of the $D(2,1;\alpha)$ family.
It turns out that the value of~$\alpha$ is determined (up to an $S_3$
automorphism) by the relation
\begin{equation} \label{alpharel}
\alpha \= -\sfrac12|4{-}d| \= -\sfrac1{2|\la_x|}\ .
\end{equation}
In fact, only for the special values 
\begin{equation}
\alpha\in\bigl\{ 
-3,-2,-\sfrac32,-1,-\sfrac23,-\sfrac12,-\sfrac13,\ 0,\ \sfrac12,\ 1,\ 2,\ \infty\bigr\}
\end{equation}
attained via~(\ref{alpharel}) (and its $S_3$ orbit) is $D(2,1;\alpha)$ extendable 
to an $\Ncal{=}8$ superconformal algebra.

\bigskip

\section{Duality and coupling in the homogeneous case}

\noindent
{}From the results of the previous section, an obvious duality relates
\begin{equation} \label{duality}
d\ \leftrightarrow\ 8{-}d \qquad\Leftrightarrow\qquad
\la_x\ \leftrightarrow\ -\la_x \qquad\Leftrightarrow\qquad
\bigl\{r\leftrightarrow\sfrac1r \and S^{d-1}\leftrightarrow S^{7-d}\bigr\}\ .
\end{equation}
The self-dual point at $d{=}4$, however, represents a degenerate case,
and the case of $d{=}0$ is empty.
We summarize the values for all dimensions in the following table, which
displays also the manifest ${\cal N}{=}4$ superalgebra ${\cal G}_4$ and the full 
${\cal N}{=}8$ superalgebra ${\cal G}_8$ for each case.
\begin{center}
\begin{tabular}{|c|ccccccccc|} \hline
$d$ & 0 & 1 & 2 & 3 & 4 & 5 & 6 & 7 & 8 \\ \hline $\vphantom{\Big|}$
$\Phi$ & $r^2$ & $r$ & $1$ & $r^{-1}$ & $r^{-2}$ & $r^{-3}$ & $r^{-4}$ & $r^{-5}$ & $r^{-6}$ \\
$\la_x$ & $-\sfrac14$ & $-\sfrac13$ & $-\sfrac12$ & $-1$ & $\infty$ & 
$+1$ & $+\sfrac12$ & $+\sfrac13$ & $+\sfrac14$ \\[6pt]
$\alpha$ & $-2$ & $-\sfrac32$ & $-1$ & $-\sfrac12$ & $0$ & 
$-\sfrac12$ & $-1$ & $-\sfrac32$ & $-2$ \\[6pt]
${\cal G}_4$ & $D(2,1)$ &  {\!\small $D(2,1;\sfrac12)$\!} &  $A(1,1)$ & 
$D(2,1)$ &  $A(1,1)$ &  $D(2,1)$ &  $A(1,1)$ & 
 {\!\small $D(2,1;\sfrac12)$\!} &  $D(2,1)$ \\[6pt]
${\cal G}_8$ & $D(4,1)$ & $F(4)$ & $A(3,1)$ & $D(2,2)$ & $-$ & 
$D(2,2)$ & $A(3,1)$ & $F(4)$ & $D(4,1)$ \\[6pt]
\hline
\end{tabular}
\end{center}

The duality map indicated in~(\ref{duality}) is easily performed by interchanging propagating
and auxiliary bosons and flipping the direction of the supersymmetry transformations.
If we summarily denote the propagating bosons, fermions and auxiliary bosons by
$x^a$, $\psi^i$ and~$f^\alpha$, respectively, and indicate the components of the dual
multiplet by overtildes and lowered indices, the structure schematically takes the following form,
\begin{center}
\begin{tabular}{ccccccccc}
& & $\qquad x^a$ & $\Qr$ & $\psi^i$ & $\Qr$ & $(f^\a,\dot{x}^a)$ & $\Qr$ & $\dot{\psi}^i$ \\
& & $\qquad\updownarrow$ & & $\updownarrow$ & & $\updownarrow\qquad$ & & \\
$\dot{\widetilde{\psi}_i}$ & $\Ql$ & $(\dot{\widetilde{x}}_a,\widetilde{f}_a)$ & $\Ql$ & 
$\widetilde{\psi}_i$ & $\Ql$ & $\widetilde{x}_\a \qquad$ & & \\
\end{tabular}
\end{center}
where the horizontal arrows encode the various supersymmetry transformations and the
vertical arrows depict the duality relations.

We have essentially three different cases of such a duality for ${\cal N}{=}8$ superconformal theories:
\begin{equation}
(1,8,7)\ \leftrightarrow\ (7,8,1) \quad,\qquad
(2,8,6)\ \leftrightarrow\ (6,8,2) \quad,\qquad
(3,8,5)\ \leftrightarrow\ (5,8,3) \ .
\end{equation}
The two members of each pair have different target dimensions but share the same
superconformal algebra. For this reason, they can be coupled together in a Lagrangian
\begin{equation}
\Lcal_{d+(8-d)} \= \Lcal_d + \Lcal_{8-d} + \gamma\,\Lcal_{d,(8-d)}\ ,
\end{equation}
with a coupling of dimensionless strength~$\g$ provided by a canonical pairing,
\begin{equation}
\Lcal_{d,(8-d)} \= x^a \widetilde{f}_a + \psi^i \widetilde{\psi}_i - f^\a \widetilde{x}_\a\ .
\end{equation}
Note that the dimensions in each pairing add up to one, and the duality guarantees 
the ${\cal N}{=}8$ superconformal invariance of the coupling term, as long as the
transformations remain homogeneous. This is the case for $d{=}1$ and $d{=}2$.
In three dimensions, there exists an inhomogeneous deformation of the (3,8,5)
multiplet. When this is turned on, the coupling to the dual (5,8,3) becomes less obvious.
We will dwell on this point later on.

Let us take a look at the bosonic part of the Lagrangian in the homogeneous case.
It takes the form
\begin{equation}
\Lcal_{d+(8-d)}\big| \= 
\Phi(r)\,\bigl(\dot{x}^a\dot{x}^a+f^\a f^\a\bigr)+
\widetilde{\Phi}(\widetilde{r})\,\bigl(
\dot{\widetilde{x}}_\a\dot{\widetilde{x}}_\a+\widetilde{f}_a\widetilde{f}_a\bigr)+
\g\,\bigl(x^a \widetilde{f}_a - f^\a \widetilde{x}_\a \bigr)\ .
\end{equation}
We may add Fayet-Iliopoulos terms with dimensionful parameters $\mu_\a$ and $\widetilde\mu^a$
to get
\begin{equation}
\Lcal_{d+(8-d)}'\big| \= \Lcal_{d+(8-d)}\big| + \mu_\a f^\a - \widetilde\mu^a \widetilde{f}_a\ .
\end{equation}
Eliminating the auxiliary components by their equations of motion,
\begin{equation}
f_\a\=\sfrac12\Phi^{-1}\bigl(\g\widetilde{x}_\a-\mu_\a\bigr) \und 
\widetilde{f}^a\=-\sfrac12\widetilde{\Phi}^{-1}\bigl(\g x^a-\widetilde{\mu}^a\bigr)\ ,
\end{equation}
we arrive at 
\begin{equation}
\Lcal_{d+(8-d)}''\big| \=
\Phi\,\dot{x}^a\dot{x}^a+\widetilde{\Phi}\,\dot{\widetilde{x}}_\a\dot{\widetilde{x}}_\a-
\sfrac14\Phi^{-1}\bigl(\g\widetilde{x}_\a{-}\mu_\a\bigr)\bigl(\g\widetilde{x}_\a{-}\mu_\a\bigr)-
\sfrac14\widetilde{\Phi}^{-1}\bigl(\g x^a{-}\widetilde{\mu}^a\bigr)\bigl(\g x^a{-}\widetilde{\mu}^a\bigr)\ ,
\end{equation}
which features a very specific potential in the joint target space of both multiplets.

For a physical interpretation, it is useful to fix $\Phi(r)=r^{2-d}$ and $\widetilde\Phi=\widetilde{r}^{d-6}$
and pass to standard radial coordinates (up to a factor of $\sfrac12$),
\begin{equation}
\rho(r) \= \sfrac{2}{|4-d|} r^{(4-d)/2} \und 
\widetilde{\rho}(\widetilde{r})\= \sfrac{2}{|4-d|} \widetilde{r}^{(d-4)/2}\ .
\end{equation}
Introducing total angular momenta $\ell$ and $\widetilde\ell$ for the $d$- and
$(8{-}d)$-dimensional targets and unit vectors via $x^a=r e^a$ and 
$\widetilde{x}_\a=\widetilde{r}\widetilde{e}_\a$, one arrives at
\begin{equation}
\Lcal_{d+(8-d)}^{\rm cone}\big| \= \dot{\rho}^2 + \sfrac{4\ell^2}{|d-4|^2}\rho^{-2} + 
\dot{\widetilde{\rho}}^2 + \sfrac{4\widetilde\ell^2}{|d-4|^2}\widetilde\rho^{-2}-
\sfrac14\Phi^{-1}\bigl(\g \widetilde{r}\vec{\widetilde{e}}-\vec{\mu}\bigr)^2-
\sfrac14\widetilde{\Phi}^{-1}\bigl(\g r\vec{e}-\vec{\widetilde{\mu}}\bigr)^2\ ,
\end{equation}
where $r=r(\rho)$ and $\widetilde{r}=\widetilde{r}(\widetilde{\rho})$ is understood.
Apart from the standard angular momentum `barriers',
the potential for the coordinates $r$ and $\widetilde{r}$ is of oscillator
type, centered around $\vec{r}=\vec{\widetilde{\mu}}/\g$ and 
$\vec{\widetilde{r}}=\vec{\mu}/\g$ and with (position-dependent) frequencies
$\omega=\sfrac{\g}2\widetilde{\Phi}^{-1/2}$ and $\widetilde{\omega}=\sfrac{\g}2\Phi^{-1/2}$, 
respectively.

\bigskip

\section{D-module representation of the (3,8,5) supermultiplet}

\noindent
Let us adopt a convenient notation for the components of the (3,8,5) multiplet:
\begin{equation}
\begin{cases}
\textrm{bosons $x^a$:} & x, y, z \quad{\rm or}\quad x_1,x_2,x_3 \\
\textrm{fermions $\psi^i$:} & \psi_0,\psi_1,\psi_2,\psi_3,\xi_0,\xi_1,\xi_2,\xi_3 \\
\textrm{auxiliaries $f^\a$:} & f_1, f_2, g, g_1, g_2  \end{cases} \ .
\end{equation}
For simplicity, we lower all indices.
According to the relations of Section~2, we have $\la_x=-1$ and $\alpha=-\sfrac12$,
so the $\Ncal{=}4$ algebra $D(2,1;-\sfrac12)\simeq D(2,1;1)\simeq osp(4|2)$ should get enlarged
to an $D(2|2)\simeq osp(4|4)$ algebra. For the conformal factor we expect $\Phi=\sfrac1r$.

A unique feature specific to $d{=}3$ is the option to deform the homogeneous superconformal
transformations by a constant shift in some transformations of fermions to auxiliaries.
Without loss of generality, we choose a frame in which only the auxiliary coordinate~$g$
appears shifted, and only in the action of $Q_2$, $Q_3$, $Q_6$ and~$Q_7$. Hence, half
of the deformation is taken to be contained in manifestly realized $\Ncal{=}4$ supersymmetry.

The $\Ncal{=}8$ transformations are captured in the following array:
\begin{eqnarray} \label{385Q}
\begin{array}{|c|c|c|c|c|c|c|c|c|}\hline
    & Q_{8} & Q_{1} & Q_{2} & Q_{3} & Q_{4} & Q_{5} & Q_{6} & Q_{7} \\\hline
    & & & & & & & & \\[-10pt]
  x_1 & \psi_{0}& \psi_{1} & \psi_{2} & \psi_{3} & \xi_{0} & \xi_{1} & \xi_{2} & \xi_{3} \\
  x_2 & \psi_{1} & -\psi_{0}& \psi_{3} & -\psi_{2} & \xi_{1} & -\xi_{0} & -\xi_{3} & \xi_{2} \\
  x_3 & \xi_{0} &-\xi_{1} & -\xi_{2} & -\xi_{3} & -\psi_{0} & \psi_{1} & \psi_{2} & \psi_{3} \\\hline
    & & & & & & & & \\[-10pt]
  \psi_{0} & \dot{x}_1 & -\dot{x}_2 & -f_{1} & -f_{2} & -\dot{x}_3 & -g & -g_{1} & -g_{2} \\
  \psi_{1} & \dot{x}_2 & \dot{x}_1 & -f_{2} & f_{1} & -g & \dot{x}_3 & g_{2} & -g_{1} \\
  \psi_{2}& f_{1} & f_{2} & \dot{x}_1 & -\dot{x}_2 & -g_{1} & -g_{2} & \dot{x}_3 & g{+}c \\
  \psi_{3} & f_{2} & -f_{1} & \dot{x}_2 & \dot{x}_1 & -g_{2} & g_{1} & -g{-}c & \dot{x}_3 \\
  \xi_{0} & \dot{x}_3 & g & g_{1} & g_{2} & \dot{x}_1 & -\dot{x}_2 & -f_{1} & -f_{2} \\
  \xi_{1} & g & -\dot{x}_3 & g_{2} & -g_{1} & \dot{x}_2 & \dot{x}_1 & f_{2} & -f_{1} \\
  \xi_{2} & g_{1} & -g_{2}  & -\dot{x}_3 & g{+}c& f_{1} & -f_{2} & \dot{x}_1 & \dot{x}_2 \\
  \xi_{3} & g_{2} & g_{1} & -g{-}c & -\dot{x}_3 & f_{2} & f_{1} & -\dot{x}_2 & \dot{x}_1 \\\hline
    & \qquad\quad & \qquad\quad & \qquad\quad & \qquad\quad & \qquad\quad & \qquad\quad & 
\qquad\quad & \qquad\quad \\[-10pt]
  f_{1} & \dot{\psi}_{2} & -\dot{\psi}_{3} & -\dot{\psi}_{0} & \dot{\psi}_{1} & \dot{\xi}_{2} & \dot{\xi}_{3} & -\dot{\xi}_{0}& -\dot{\xi}_{1} \\
  f_{2} & \dot{\psi}_{3} & \dot{\psi}_{2} & -\dot{\psi}_{1} & -\dot{\psi}_{0} & \dot{\xi}_{3} & -\dot{\xi}_{2}& \dot{\xi}_{1}& -\dot{\xi}_{0} \\
  g & \dot{\xi}_{1} & \dot{\xi}_{0} & -\dot{\xi}_{3} & \dot{\xi}_{2} & -\dot{\psi}_{1}& -\dot{\psi}_{0} & -\dot{\psi}_{3}& \dot{\psi}_{2} \\
  g_{1} & \dot{\xi}_{2} & \dot{\xi}_{3} & \dot{\xi}_{0} & -\dot{\xi}_{1} &-\dot{\psi}_{2} & \dot{\psi}_{3} & -\dot{\psi}_{0} & -\dot{\psi}_{1}\\
  g_{2} & \dot{\xi}_{3} & -\dot{\xi}_{2} & \dot{\xi}_{1}& \dot{\xi}_{0} & -\dot{\psi}_{3} & -\dot{\psi}_{2}& \dot{\psi}_{1}&-\dot{\psi}_{0}\\\hline
\end{array}
\end{eqnarray}

The action for the (3,8,5) multiplet reads
\begin{equation}
S_3 \= \int\!\diff t\ \Lcal_3 \= 
\int\!\diff t\ Q_8 Q_1 Q_2 Q_3 F(x,y,z)
\end{equation}
with 
\begin{equation}
F_{xx}+F_{yy}+F_{zz} \= 0\ ,
\end{equation}
and the conformal factor comes out as
\begin{equation}
F_{zz} \= \Phi \= \sfrac1r 
\quad\with\quad r^2=x^2+y^2+z^2\ .
\end{equation}
Without loss of generality, the $z$ coordinate is singled out because 
we had to make a choice in the supersymmetry transformations.

The dependence on the inhomogeneous shift parameter~$c$ is linear, so we write
\begin{equation}
{\cal L}_3 \= {\cal L}_3^{(0)} + c\,{\cal L}_3^{(1)}\ .
\end{equation}
After a lengthy but straightforward computation, we find
\begin{equation}
{\cal L}_3^{(0)} \=
\Phi\,(\dot{x}^{2}+\dot{y}^{2}+\dot{z}^{2}+f_{1}^{2}+f_{2}^{2}+g^{2}+g_{1}^{2}+g_{2}^{2})
\ +\ {\rm fermionic \ terms}
\end{equation}
and
\begin{eqnarray}
{\cal L}_3^{(1)} &=&\Phi\,g+A_x\dot{x}+A_y\dot{y}+\\\nonumber
&&\Phi_{x}(\psi_{0}\xi_{1}+\psi_{1}\xi_{0})+\Phi_{y}(\psi_{1}\xi_{1}-\psi_{0}\xi_{0})
-\Phi_{z}(\psi_{1}\psi_{0}+\xi_{1}\xi_{0})\ ,
\end{eqnarray}
where we introduced
\begin{equation}
A_x \= F_{zy} \und A_y \= -F_{zx}\ .
\end{equation}
The complete expression of ${\cal L}_3^{(0)}$ is displayed in Appendix~A.
Setting all fermions to zero, we extract the bosonic part
\begin{equation}
{\cal L}_3\big| \= \Phi\, (\dot{x}^{2}+\dot{y}^{2}+\dot{z}^{2}+f_{1}^{2}+f_{2}^{2}+g^{2}+g_{1}^{2}+g_{2}^{2})
\ +\  c\,(\Phi\,g +A_x\dot{x}+A_y\dot{y})\ .
\end{equation}
We remark that only the $z$~derivative of~$F$ appears, so it makes sense to define a prepotential
\begin{equation}
G\ :=\ F_z \qquad\Rightarrow\qquad
G_x \= -A_y\ ,\quad G_y \= A_x\ ,\quad G_z \= \Phi\ ,
\end{equation}
which inherits the harmonicity from~$F$.

It is admissible to slightly deform our model by adding Fayet-Iliopoulos terms.
This extends the bosonic Lagrangian to
\begin{equation}
{\cal L}'_3\big|\= {\cal L}_3\big|+ \mu_\a f_\a + \zeta g +\zeta_\a g_\a
\quad\with \a=1,2
\end{equation}
and five real parameters $\mu_\a$, $\zeta$ and $\zeta_\a$.

We solve the equations of motion for the auxiliary fields,
\begin{equation}
f_\a\=-\frac{\mu_{\a}}{2\Phi}\ ,\qquad
g\=-\frac{\zeta+c\Phi}{2\Phi}\ ,\qquad
g_{\a}\=-\frac{\zeta_{\a}}{2\Phi}\ ,
\end{equation}
and eliminate them from the Lagrangian to arrive at
\begin{equation}
{\cal L}''_3\big|\=\Phi\,
(\dot{x}^{2}+\dot{y}^{2}+\dot{z}^{2})\ -\
\sfrac14\Phi^{-1}(\mu_\a^2+\zeta^2+\zeta_\a^2)\ -\
\sfrac12 c\,\zeta\ -\ \sfrac14 c^{2}\Phi\ +\
c\,(A_x\dot{x}+A_y\dot{y})\ .
\end{equation} 
Apparently, there is not only a magnetic but also an electric field, together
\begin{equation}
B_x=c\,G_{xz}\ ,\quad B_y=c\,G_{yz}\ ,\quad 
B_z=-c (G_{xx}{+}G_{yy})=c\,G_{zz}
\und E_a = -\sfrac14 c^2 G_{za}\ ,
\end{equation}
both being simply proportional to the gradient of $G_z=\Phi$.
With $\Phi=\sfrac1r$, we identify a magnetic monopole,
while for the interpretation of the electric field we better pass
to the conical coordinates,
\begin{equation}
r=\sfrac14\rho^2 \qquad\Rightarrow\qquad
\diff s^2 \=\diff\rho^2+\sfrac14\rho^2\diff\Omega_2^2 \und
A_0\=c^2\rho^{-2}\ .
\end{equation}
The bosonic dynamics of this theory has been analyzed for general
values of~$\alpha$ in~\cite{IKL}.

\bigskip

\section{Gauge freedom}

\noindent
In order to explicitly write down the Lagrangian, we must `integrate' $\Phi$ 
to find the prepotential~$G$, from which the gauge potential~$A$ is obtained.
The answer is not unique, due to abelian gauge invariance,
\begin{equation}
\de A_a \= \pa_a u \und \de G \= v
\end{equation}
with a priori arbitrary harmonic gauge functions $u$ and~$v$.
However, the invariance of $\Phi{=}G_z$ enforces $v_z=0$, and
the relation between $G$ and $A_a$ connects the two functions,
\begin{equation}
u_x = v_y \und u_y = -v_x\ .
\end{equation}
The (local) solution introduces another function~$h(x,y)$ via
\begin{equation}
u = h_y(x,y) + \widetilde{u}(z) \und v = h_x(x,y)
\quad\with\quad h_{xx}+h_{yy}=0\ .
\end{equation}
The harmonicity of $u$ implies that $\widetilde{u}$ is at most linear in~$z$.
Alternatively, we may interpret the above relation as Cauchy-Riemann equations
for the real and imaginary part of a holomorphic function of \ $w=x+\im y$,
\begin{equation}
v-\im (u{-}\widetilde{u}) \= E(w) \ =:\ \pa_w H(w) \qquad\Rightarrow\qquad
h \= H(w) + \overline{H}(\overline{w})\ ,
\end{equation}
where $H$ and $h$ are determined up to a constant.
Therefore, the gauge freedom for the prepotential is encoded in a single
holomorphic function~$E$.

For the magnetic monopole there does not exist a globally regular gauge potential;
we must be content with configurations on a `northern' (N) and on a `southern' (S) patch,
related by a regular gauge transformation in the equatorial overlap. The standard
expressions obtained from $G_z=\sfrac1r$ read
\begin{eqnarray}
G^N \= +\ln(r{+}z) \qquad&\Rightarrow&\qquad 
A^N_x=G^N_y=\sfrac{y}{r(z{+}r)} \und A^N_y=-G^N_x=-\sfrac{x}{r(z{+}r)} \ , \\
G^S \,\= -\ln(r{-}z) \qquad&\Rightarrow&\qquad 
A^S_x\,=G^S_y\,=\sfrac{y}{r(z{-}r)} \und A^S_y\,=-G^S_x\,=-\sfrac{x}{r(z{-}r)} \ , 
\end{eqnarray}
so that indeed (for $a=x,y$)
\begin{equation}
G^N-G^S\= \ln(x^2{+}y^2)\ =:\ h_x  \qquad\Rightarrow\qquad
A^N_a-A^S_a\=-2\pa_a\arctan\sfrac{y}{x}\ =:\ \pa_a h_y\ ,
\end{equation}
and the holomorphic combination
\begin{equation}
E\=\ln(x^2{+}y^2)+2\im\arctan\sfrac{y}{x}\=\ln w^2
\end{equation}
gives rise to the correct `pre-gauge' function
\begin{equation}
H\=2w(\ln w{-}1) \qquad\Rightarrow\qquad
h\=x\ln(x^2{+}y^2)-2x-2y\arctan\sfrac{y}{x}
\end{equation}
in the class described above and regular away from the poles.
The singularity of the northern functions along the negative $z$-axis and likewise
for the southern patch signify the would-be Dirac string in a global configuration.

\bigskip

\section{The dual (5,8,3) supermultiplet}

\noindent
Applying the duality reflection to the (3,8,5)~multiplet, we obtain a (5,8,3) multiplet.
However, we must first put the inhomogeneous deformation parameter~$c$ to zero, since
such a deformation does not exist for $d{=}5$.
Section~2 tells us that $\la_x=+1$ and $\Phi=r^{-3}$, and we again realize an 
$D(2,2)\simeq osp(4|4)$ superalgebra. 
Naming the components as follows,
\begin{equation}
\begin{cases}
\textrm{bosons $\tilde{x}_\a$:} & v_1,v_2,w,w_1,w_2 \\
\textrm{fermions $\tilde{\psi}_i$:} & \chi_0,\chi_1,\chi_2,\chi_3,\la_0,\la_1,\la_2,\la_3 \\
\textrm{auxiliaries $\tilde{f}_a$:} & h_1,h_2,h_3  \end{cases} \ ,
\end{equation}
the array~(\ref{385Q}) gets transformed into the ${\cal N}{=}8$ transformations for the
(5,8,3) multiplet:
\begin{eqnarray} \label{583Q}
\begin{array}{|c|c|c|c|c|c|c|c|c|}\hline
    & Q_{8} & Q_{1} & Q_{2} & Q_{3} & Q_{4} & Q_{5} & Q_{6} & Q_{7} \\\hline
    & & & & & & & & \\[-10pt]
  v_1 & \la_2 & -\la_3 & -\la_0 & \la_1 & \chi_2 & \chi_3 & -\chi_0 & -\chi_1 \\
  v_2 & \la_3 & \la_2 & -\la_1 & -\la_0 & \chi_3 & -\chi_2 & \chi_1 & -\chi_0\\
  w & \chi_1 & \chi_0 & -\chi_3 & \chi_2 & -\la_1 & -\la_0 & -\la_3 & \la_2 \\
  w_1 & \chi_2 & \chi_3 & \chi_0 & -\chi_1 & -\la_2 & \la_3 & -\la_0 & -\la_1 \\
  w_2 & \chi_3 & -\chi_2 & \chi_1 & \chi_0 & -\la_3 & -\la_2 & \la_1 & -\la_0 \\\hline
    & & & & & & & & \\[-10pt]
  \chi_{0} & h_3 & \dot{w} & \dot{w}_1 & \dot{w}_2 & h_1 & -h_2 & -\dot{v}_1 & -\dot{v}_2 \\
  \chi_{1} & \dot{w} & -h_3 & \dot{w}_2 & -\dot{w}_1 & h_2 & h_1 & \dot{v}_2 & -\dot{v}_1\\
  \chi_{2} & \dot{w}_1 & -\dot{w}_2 & -h_3 & \dot{w} & \dot{v}_1 & -\dot{v}_2 & h_1 & h_2 \\
  \chi_{3} & \dot{w}_2 & \dot{w}_1 & -\dot{w} & -h_3 & \dot{v}_2 & \dot{v}_1 & -h_2 & h_1 \\
  \la_{0} & h_1 & -h_2 & -\dot{v}_1 & -\dot{v}_2 & -h_3 & -\dot{w} & -\dot{w}_1 & -\dot{w}_2 \\
  \la_{1} & h_2 & h_1 & -\dot{v}_2 & \dot{v}_1 & -\dot{w} & h_3 & \dot{w}_2 & -\dot{w}_1 \\
  \la_{2} & \dot{v}_1 & \dot{v}_2 & h_1 & -h_2 & -\dot{w}_1 & -\dot{w}_2 & h_3 & \dot{w} \\
  \la_{3} & \dot{v}_2 & -\dot{v}_1 & h_2 & h_1 & -\dot{w}_2 & \dot{w}_1 & -\dot{w} & h_3 \\\hline
    & \qquad\quad & \qquad\quad & \qquad\quad & \qquad\quad & \qquad\quad & \qquad\quad & 
\qquad\quad & \qquad\quad \\[-10pt]
  h_1 & \dot{\la}_0 & \dot{\la}_1 & \dot{\la}_2 & \dot{\la}_3 & \dot{\chi}_0 & \dot{\chi}_1 & \dot{\chi}_2 & \dot{\chi}_3 \\
  h_2 & \dot{\la}_1 & -\dot{\la}_0 & \dot{\la}_3 & -\dot{\la}_2 & \dot{\chi}_1 & -\dot{\chi}_0 & -\dot{\chi}_3 & \dot{\chi}_2 \\
  h_3 & \dot{\chi}_0 & -\dot{\chi}_1 & -\dot{\chi}_2 & -\dot{\chi}_3 & -\dot{\la}_0 & \dot{\la}_1 & \dot{\la}_2 & \dot{\la}_3 \\\hline
\end{array}
\end{eqnarray}
The full Lagrangian $\Lcal_5$ is found in Appendix~B. Its bosonic part is obvious,
\begin{equation}
\Lcal_5\big|\=
\widetilde{\Phi}\,(\dot{v}_\alpha^{2}+\dot{w}^{2}+\dot{w}_\alpha^{2}+h_{a}^{2})\ ,
\end{equation}
where the prepotential function is
\begin{equation}
\widetilde\Phi\=F_{v_{1}v_{1}}+F_{v_{2}\upsilon_{2}}\=-(F_{ww}+F_{w_{1}w_{1}}+F_{w_{2}w_{2}})\ .
\end{equation}

\bigskip

\section{Coupling (3,8,5) to (5,8,3)}

\noindent
Since both (3,8,5) and (5,8,3) multiplets represent the same $D(2,2)$ superalgebra, 
it is natural to couple them. The duality provides a canonical interaction term 
$\Lcal_{3,5}^{(0)}$ in the joint Lagrangian
\begin{equation}
\Lcal_3^{(0)}\ +\ \Lcal_5\ +\ \gamma\Lcal_{3,5}^{(0)}
\end{equation}
of the form
\begin{equation} \label{pairing}
\Lcal_{3,5}^{(0)} \= x_a h_a - f_\a v_\a - g\,w - g_\a w_\a + \psi_i\la_i +  \xi_i\chi_i 
\quad\with a=1,2,3\ ,\quad \a=1,2\ ,\quad i=0,1,2,3\ ,
\end{equation}
with some dimensionless coupling constant~$\gamma$.
It is easy to check that $\Lcal_{3,5}^{(0)}$ is invariant (up to total time derivatives) under
all eight supersymmetries and their conformal partners, because the dimensions of any two
duality partners add up to one.

The superscript~$(0)$ reminds us that we turned off the inhomogeneous deformation in the
(3,8,5)~multiplet. So the question arises as to whether it is possible to extend this coupling
to the deformed multiplet as well, and what this entails for the dual (5,8,3)~multiplet.
To answer this, we first observe that 
\begin{equation}
\Lcal_3\ +\ \Lcal_5\ +\ \gamma\Lcal_{3,5}^{(0)}
\end{equation}
is indeed invariant (up to total time derivatives) under $Q_8$, $Q_1$, $Q_4$ and~$Q_5$, but
\begin{equation} \label{extra}
Q_2\Lcal_{3,5}^{(0)}=-c\chi_3\ ,\quad
Q_3\Lcal_{3,5}^{(0)}=  c\chi_2\ ,\quad
Q_6\Lcal_{3,5}^{(0)}=-c\la_3\ ,\quad
Q_7\Lcal_{3,5}^{(0)}=  c\la_2
\end{equation}
do not vanish. Yet, since $c$ is a constant, these terms are linear and may be cancelled
by adding other linear terms to the interaction. To achieve this feat, however, one must
view the deformation parameter~$c$ as the highest component of an ${\cal N}{=}4$ multiplet 
of type (3,4,1) involving the supercharges $Q_j$ for $j=2,3,6,7$. 
Denoting the components of dimension $-1$, $-\sfrac12$ and $0$ by
$e_a$, $\o_i$ and $c$, respectively, the transformation table takes the form
\begin{eqnarray} \label{341Q}
\begin{array}{|c|c|c|c|c|c|c|c|c|}\hline
  & Q_{8} & Q_{1} & Q_{2} & Q_{3} & Q_{4} & Q_{5} & Q_{6} & Q_{7} \\\hline
  & & & & & & & & \\[-10pt]
  e_1 & 0 & 0 & \o_2 & \o_3 & 0 & 0 & \o_0 & \o_1 \\
  e_2 & 0 & 0 & \o_3 & -\o_2 & 0 & 0 & -\o_1 & \o_0 \\
  e_3 & 0 & 0 & -\o_0 & -\o_1 & 0 & 0 & \o_2 & \o_3 \\\hline
  & & & & & & & & \\[-10pt]
  \o_0 & 0 & 0 & 0 & -c & 0 & 0 & 0 & 0 \\
  \o_1 & 0 & 0 & c & 0 & 0 & 0 & 0 & 0 \\
  \o_2 & 0 & 0 & 0 & 0 & 0 & 0 & 0 & -c \\
  \o_3 & 0 & 0 & 0 & 0 & 0 & 0 & c & 0 \\\hline
   & \qquad\quad & \qquad\quad & \qquad\quad & \qquad\quad
   & \qquad\quad & \qquad\quad & \qquad\quad & \qquad\quad \\[-10pt]
  c & 0 & 0 & 0 & 0 & 0 & 0 & 0 & 0 \\\hline
\end{array}
\end{eqnarray}
It is important to realize that all these components are constants, i.e.~time independent,
otherwise there could not be zeros in this table. For the same reason, it is admissible
to have this multiplet annihilated by the other four supercharges, $Q_k$ for $k=8,1,4,5$.
If we add to our interaction Lagrangian two extra pieces,
\begin{equation}
\Lcal_{3,5}^{(1)}\= \o_0\chi_2+\o_1\chi_3+\o_2\la_2+\o_3\la_3 \und 
\Lcal_{3,5}^{(2)}\= e_1 h_1 + e_2 h_2 + e_3 h_3 \ ,
\end{equation}
it is not hard to check that all unwanted terms get cancelled, and only total time derivatives
remain. In other words,
\begin{equation} \label{Ltotal}
\Lcal_{3+5} \ :=\ \Lcal_3\ +\ \Lcal_5\ +\ \gamma\Lcal_{3,5}
\end{equation}
is fully ${\cal N}{=}8$ superconformally invariant for
\begin{eqnarray}
\Lcal_{3,5} &=& (x_a{+}e_a)h_a - f_\a v_\a - g\,w - g_\a w_\a \\
&+&\xi_0\chi_0+\xi_1\chi_1+(\xi_2{+}\omega_0)\chi_2+(\xi_3{+}\omega_1)\chi_3+
\psi_0\la_0+\psi_1\la_1+(\psi_2{+}\omega_2)\la_2+(\psi_3{+}\omega_3)\la_3\ ,\nonumber
\end{eqnarray}
which adds to the pairings~(\ref{pairing}) a term linear in a (1,4,3) submultiplet
$(w;\chi_2,\chi_3,\la_2,\la_3;h_a)$ inside our dual (5,8,3) multiplet.
Another interpretation is that the (1,8,5) components with inhomogeneous transformation
receive constant shifts which cancel the inhomogeneity produced in the canonical coupling term.

Interestingly, there is another way to cancel the non-invariant terms~(\ref{extra}).
Observing that
\begin{equation} \label{extra2}
Q_j \Lcal_{3,5}^{(0)} \= c\,Q_j w \quad\for j=2,3,6,7
\end{equation}
suggests repairing the deficit by adding 
\begin{equation}
\Lcal_{3,5}^{(0')}\= - c\,w
\end{equation}
to the interaction. While $Q_j\Lcal_{3,5}^{(0')}$ indeed just cancels the unwanted terms,
now the other four supersymmetries are compromised, however, as
\begin{equation}
Q_8\Lcal_{3,5}^{(0')}=-c\,\chi_1\ ,\quad
Q_1\Lcal_{3,5}^{(0')}=-c\,\chi_0\ ,\quad
Q_4\Lcal_{3,5}^{(0')}=c\,\la_1\ ,\quad
Q_5\Lcal_{3,5}^{(0')}=c\,\la_0\ .
\end{equation}
Comparing with~(\ref{extra}), we see that the deficiency has simply been shifted from
the $Q_j$ to the $Q_k$ with $k=8,1,4,5$, and the relevant fermionic components carry indices
0 and 1 instead of 2 and~3. Hence, adding a suitable constant (3,4,1) multiplet for those
supersymmetries and the appropriate terms $\Lcal_{3,5}^{(1')}$ and $\Lcal_{3,5}^{(2')}$
to the interaction will accomplish the job just as well.
The only difference for the bosonic Lagrangians is an additional term of~$-\g c\,w$.

Sticking with the first resolution and adding Fayet-Iliopoulos terms for all auxiliary components, 
the bosonic part of the total action reads
\begin{eqnarray}
\Lcal_{3+5}'\big| &=&
\Phi\,\bigl(\dot{x}_a^2+f_\a^2+g^2+g_\a^2\bigr)+c\,\vec{A}{\cdot}\dot{\vec{x}}+
\widetilde{\Phi}\,\bigl(\dot{v}_\a^2+\dot{w}^2+\dot{w}_\a^2+h_a^2\bigr) \nonumber \\
&-& \bigl(\g v_\a{-}\mu_\a\bigr)\,f_\a-\bigl(\g w{-}\zeta{-}c\Phi\bigr)\,g-
\bigl(\g w_\a{-}\zeta_\a\bigr)\,g_\a+\bigl(\g (x_a{+}e_a){-}\widetilde{\mu}_a\bigr)\,h_a\ ,
\end{eqnarray}
and elimination of the auxiliary components produces
\begin{eqnarray}
\Lcal_{3+5}''\big| &=&
\Phi\,\dot{x}_a^2+c\,\vec{A}{\cdot}\dot{\vec{x}}+
\widetilde{\Phi}\,\bigl(\dot{v}_\a^2+\dot{w}^2+\dot{w}_\a^2\bigr) \nonumber \\
&-&\sfrac14\Phi^{-1}\bigl( (\g v_\a{-}\mu_\a)^2+(\g w{-}\zeta{-}c\Phi)^2+(\g w_\a{-}\zeta_\a)^2\bigr)-
\sfrac14\widetilde{\Phi}^{-1}\bigl(\g(x_a{+}e_a)-\widetilde{\mu}_a\bigr)^2\ .
\end{eqnarray}
The constant Lagrange multipliers $e_a$ serve to eliminate the zero modes of the~$h_a$.
For convenience, we relabel $w_\a=v_{2+\a}$ and $w=v_5$ and define
$v^2=v_\a v_\a+w^2+w_\a w_\a$.
In conical radial coordinates $\rho=2r^{1/2}$ and $\sigma=2v^{-1/2}$,
the bosonic action then takes the form
\begin{eqnarray}
\Lcal_{3+5}^{\rm cone}\big| &=&
\dot{\rho}^2+4\ell^2\rho^{-2}+\dot{\sigma}^2+4\widetilde{\ell}^2\sigma^{-2} 
+c\,\vec{A}{\cdot}\dot{\vec{x}} \nonumber\\
&-&\sfrac1{16}\rho^2\bigl(4\g\sigma^{-2}\vec{e}_\sigma-\vec{\mu}-4c\rho^{-2}\vec{e}_5\bigr)^2
-\sigma^{-6}\bigl(\g(\rho^2\vec{e}_\rho{+}4\vec{e})-4\vec{\widetilde{\mu}}\bigr)^2\ ,
\end{eqnarray}
where we introduced the angular momenta $\ell$ and $\widetilde{\ell}$ 
in the three- and five-dimensional targets, and the vectors in the first and second brackets are
five- and three-dimensional, respectively.

\bigskip

\section{A deformed (5,8,3) supermultiplet}

\noindent
If in (\ref{Ltotal}) we set to zero the complete (5,8,3) multiplet, we simply come back to 
the original deformed (3,8,5) theory. Let us then try the opposite and see whether we recover
the (5,8,3)~model. However, due to~(\ref{extra}) it is not consistent to put the (3,8,5)
components to zero completely, but we must keep the zero modes of $x_a$, $\psi_2$, $\psi_3$,
$\xi_2$, $\xi_3$ and~$g$, which we denote by an overbar. With this provision, the full Lagrangian
(\ref{Ltotal}) reduces to
\begin{eqnarray}
\widehat{\Lcal}_5 &=&\Lcal_5\ +\ \gamma\,\bigl( 
(\overline{x}_a{+}e_a) h_a +
(\overline{\xi}_2{+}\omega_0)\chi_2 + (\overline{\xi}_3{+}\omega_1)\chi_3+
(\overline{\psi}_2{+}\omega_2)\la_2 + (\overline{\psi}_3{+}\omega_3)\la_3-
(\overline{g}{+}c) w \bigr)  \nonumber \\
&=:& \Lcal_5\ +\ \gamma\,\bigl( 
e'_a h_a+\omega'_0\chi_2+\omega'_1\chi_3+\omega'_2\la_2+\omega'_3\la_3-c'w\bigr)\ ,
\end{eqnarray} 
and to the transformations~(\ref{341Q}) of the constants we must add
\begin{eqnarray} \label{340Q}
\begin{array}{|c|c|c|c|c|c|c|c|c|}\hline
  & Q_{8} & Q_{1} & Q_{2} & Q_{3} & Q_{4} & Q_{5} & Q_{6} & Q_{7} \\\hline
  & \qquad\quad & \qquad\quad & \qquad\quad & \qquad\quad
  & \qquad\quad & \qquad\quad & \qquad\quad & \qquad\quad \\[-10pt]
  \overline{x}_1 & 0 & 0 & \overline{\psi}_{2} & \overline{\psi}_{3} 
  & 0 & 0 & \overline{\xi}_{2} & \overline{\xi}_{3} \\
  \overline{x}_2 & 0 & 0 & \overline{\psi}_{3} & -\overline{\psi}_{2} 
  & 0 & 0 & -\overline{\xi}_{3} & \overline{\xi}_{2} \\
  \overline{x}_3 & 0 & 0 & -\overline{\xi}_{2} & -\overline{\xi}_{3} 
  & 0 & 0 & \overline{\psi}_{2} & \overline{\psi}_{3} \\\hline
    & & & & & & & & \\[-10pt]
  \overline{\xi}_2 & 0 & 0 & 0 & \overline{g}{+}c & 0 & 0 & 0 & 0 \\
  \overline{\xi}_3 & 0 & 0 & -\overline{g}{-}c & 0 & 0 & 0 & 0 & 0 \\
  \overline{\psi}_2 & 0 & 0 & 0 & 0 & 0 & 0 & 0 & \overline{g}{+}c \\
  \overline{\psi}_3 & 0 & 0 & 0 & 0 & 0 & 0 & -\overline{g}{-}c & 0 \\\hline
  \overline{g} & 0 & 0 & 0 & 0 & 0 & 0 & 0 & 0 \\\hline
\end{array}
\end{eqnarray}
which is what remains of~(\ref{385Q}).
We see that only the four $Q_j$ are effective.
The upshot is a deformation of the original (5,8,3) Lagrangian by linear terms
in a (1,4,3) submultiplet. The linear coefficients $(e'_a,\omega'_i,c')$ are just 
the sum of the (3,4,1) zero-mode submultiplet~(\ref{340Q}) of the original 
(3,8,5) multiplet and the constant auxiliary (3,4,1) multiplet~(\ref{341Q}).
This combination transforms as follows,
\begin{eqnarray} 
\begin{array}{|c|c|c|c|c|c|c|c|c|}\hline
  & Q_{8} & Q_{1} & Q_{2} & Q_{3} & Q_{4} & Q_{5} & Q_{6} & Q_{7} \\\hline
  & & & & & & & & \\[-10pt]
  e'_1 & 0 & 0 & \o'_2 & \o'_3 & 0 & 0 & \o'_0 & \o'_1 \\
  e'_2 & 0 & 0 & \o'_3 & -\o'_2 & 0 & 0 & -\o'_1 & \o'_0 \\
  e'_3 & 0 & 0 & -\o'_0 & -\o'_1 & 0 & 0 & \o'_2 & \o'_3 \\\hline
  & & & & & & & & \\[-10pt]
  \o'_0 & 0 & 0 & 0 & -c' & 0 & 0 & 0 & 0 \\
  \o'_1 & 0 & 0 & c' & 0 & 0 & 0 & 0 & 0 \\
  \o'_2 & 0 & 0 & 0 & 0 & 0 & 0 & 0 & -c' \\
  \o'_3 & 0 & 0 & 0 & 0 & 0 & 0 & c' & 0 \\\hline
   & \qquad\quad & \qquad\quad & \qquad\quad & \qquad\quad
   & \qquad\quad & \qquad\quad & \qquad\quad & \qquad\quad \\[-10pt]
  c' & 0 & 0 & 0 & 0 & 0 & 0 & 0 & 0 \\\hline
\end{array}
\end{eqnarray}
Hence, the coupling of the (5,8,3) multiplet to a dual inhomogeneous (3,8,5) multiplet
leads to a deformation of the former, which consists of the coupling of a (1,4,3) submultiplet
to an auxiliary constant (3,4,1) dual multiplet. The deformation is parametrized by~$\g$
and contains the (3,8,5) inhomogeneity~$c$ as part of it.
Of course, we may also add standard Fayet-Iliopoulos terms.

\bigskip

\appendix

\section{Appendix: Action for the (3,8,5) supermultiplet}

\noindent
The complete Lagrangian for the (3,8,5) multiplet reads
\begin{eqnarray}
{\cal L}_3^{(0)} &=&
      \Phi\,(\dot{x}^{2}+\dot{y}^{2}+\dot{z}^{2}+f_{1}^{2}+f_{2}^{2}+g^{2}+g_{1}^{2}+g_{2}^{2})+\\\nonumber
       & & \Phi\,(\dot{\psi}_{0}\psi_{0}+\dot{\psi}_{1}\psi_{1}+\dot{\psi}_{2}\psi_{2}+\dot{\psi}_{3}\psi_{3}+
       \dot{\xi_{0}}\xi_{0}+\dot{\xi}_{1}\xi_{1}+\dot{\xi}_{2}\xi_{2}+\dot{\xi}_{3}\xi_{3})+\\\nonumber
       & &\Phi_{x}((\dot{y}\xi_{0}\xi_{1}+f_{1}\xi_{0}\xi_{2}+f_{2}\xi_{0}\xi_{3})+
       (\dot{y}\psi_{0}\psi_{1}+f_{1}\psi_{0}\psi_{2}+f_{2}\psi_{0}\psi_{3}) \\\nonumber
       &&+(g\psi_{0}\xi_{1}+g_{1}\psi_{0}\xi_{2}+g_{2}\psi_{0}\xi_{3})+(g\psi_{1}\xi_{0}+g_{1}\psi_{2}\xi_{0}+g_{2}\psi_{3}\xi_{0})\\\nonumber
       &&-(\dot{y}\xi_{2}\xi_{3}+f_{1}\xi_{3}\xi_{1}+f_{2}\xi_{1}\xi_{2})+(\dot{y}\psi_{2}\psi_{3}+f_{1}\psi_{3}\psi_{1}+f_{2}\psi_{1}\psi_{2})\\\nonumber
       &&+(g\xi_{3}\psi_{2}+g_{1}\xi_{1}\psi_{3}+g_{2}\xi_{2}\psi_{1})-(g\xi_{2}\psi_{3}+g_{1}\xi_{3}\psi_{1}+g_{2}\xi_{1}\psi_{2})\\\nonumber
       &&+\dot{z}(\psi_{0}\xi_{0}+\xi_{1}\psi_{1}+\xi_{2}\psi_{2}+\xi_{3}\psi_{3}))+\\\nonumber
       &&\Phi_{y}((-\dot{x}\xi_{0}\xi_{1}-f_{1}\xi_{0}\xi_{3}+f_{2}\xi_{0}\xi_{2})+
       (\dot{x}\psi_{1}\psi_{0}-f_{1}\psi_{3}\psi_{0}+f_{2}\psi_{2}\psi_{0})\\\nonumber
       &&-(\dot{z}\xi_{1}\psi_{0}-g_{1}\xi_{3}\psi_{0}+g_{2}\xi_{2}\psi_{0})-
       (\dot{z}\xi_{0}\psi_{1}+g_{1}\xi_{0}\psi_{3}-g_{2}\xi_{0}\psi_{2})\\\nonumber
       &&+g(\xi_{0}\psi_{0}-\xi_{1}\psi_{1}+\xi_{2}\psi_{2}+\xi_{3}\psi_{3})\\\nonumber
       &&+\dot{x}(\xi_{2}\xi_{3}-\psi_{2}\psi_{3})-\dot{z}(\xi_{3}\psi_{2}-\xi_{2}\psi_{3})\\\nonumber
       &&+g_{1}(\psi_{1}\xi_{2}-\xi_{1}\psi_{2})+g_{2}(\psi_{1}\xi_{3}-\xi_{1}\psi_{3}))+\\\nonumber
       &&\Phi_{z}((g\xi_{0}\xi_{1}+g_{1}\xi_{0}\xi_{2}+g_{2}\xi_{0}\xi_{3})+
       (g\psi_{0}\psi_{1}+g_{1}\psi_{0}\psi_{2}+g_{2}\psi_{0}\psi_{3})\\\nonumber
       &&-(\dot{y}\psi_{0}\xi_{1}+f_{1}\psi_{0}\xi_{2}+f_{2}\psi_{0}\xi_{3})-(\dot{y}\psi_{1}\xi_{0}+f_{1}\psi_{2}\xi_{0}+f_{2}\psi_{3}\xi_{0}) \\\nonumber
       &&+(g\xi_{2}\xi_{3}+g_{1}\xi_{3}\xi_{1}+g_{2}\xi_{1}\xi_{2})+(g\psi_{2}\psi_{3}+g_{1}\psi_{3}\psi_{1}+g_{2}\psi_{1}\psi_{2})\\\nonumber
       &&+(\dot{y}\xi_{3}\psi_{2}+f_{1}\xi_{1}\psi_{3}+f_{2}\xi_{2}\psi_{1})-(\dot{y}\xi_{2}\psi_{3}+f_{1}\xi_{3}\psi_{1}+f_{2}\xi_{1}\psi_{2})\\\nonumber
       &&+\dot{x}(\psi_{0}\xi_{0}+\xi_{1}\psi_{1}+\xi_{2}\psi_{2}+\xi_{3}\psi_{3}))+\\\nonumber
       &&\Phi_{xx}(\psi_{3}\psi_{1}\xi_{2}\xi_{0}+\psi_{3}\psi_{0}\xi_{2}\xi_{1}-\psi_{2}\psi_{1}\xi_{3}\xi_{0}-\psi_{2}\psi_{0}\xi_{3}\xi_{1})+\\\nonumber
       &&\Phi_{yy}(\psi_{2}\psi_{0}\xi_{2}\xi_{0}+\psi_{3}\psi_{0}\xi_{3}\xi_{0}-\psi_{2}\psi_{1}\xi_{2}\xi_{1}-\psi_{3}\psi_{1}\xi_{3}\xi_{1})+\\\nonumber
       &&\Phi_{zz}(-\xi_{3}\xi_{2}\xi_{1}\xi_{0}+\psi_{3}\psi_{2}\xi_{1}\xi_{0}-\psi_{1}\psi_{0}\xi_{3}\xi_{2}+\psi_{3}\psi_{2}\psi_{1}\psi_{0})+\\\nonumber
       &&\Phi_{xy}(\psi_{2}\psi_{0}\xi_{3}\xi_{0}-\psi_{2}\psi_{0}\xi_{2}\xi_{1}+\psi_{3}\psi_{1}\xi_{2}\xi_{1}-\psi_{3}\psi_{0}\xi_{2}\xi_{0}\\\nonumber
       &&-\psi_{2}\psi_{1}\xi_{3}\xi_{1}-\psi_{3}\psi_{0}\xi_{3}\xi_{1}-\psi_{2}\psi_{1}\xi_{2}\xi_{0}-\psi_{3}\psi_{1}\xi_{3}\xi_{0})-\\\nonumber
       &&\Phi_{xz}(\psi_{2}\xi_{3}\xi_{1}\xi_{0}+\psi_{2}\psi_{1}\psi_{0}\xi_{3}-\psi_{3}\psi_{1}\psi_{0}\xi_{2}+\psi_{3}\psi_{2}\psi_{1}\xi_{0}\\\nonumber
       &&+\psi_{3}\xi_{2}\xi_{1}\xi_{0}-\psi_{0}\xi_{3}\xi_{2}\xi_{1}+\psi_{3}\psi_{2}\psi_{0}\xi_{1}-\psi_{1}\xi_{3}\xi_{2}\xi_{0})-\\\nonumber
       &&\Phi_{yz}(\psi_{0}\xi_{3}\xi_{2}\xi_{0}+\psi_{2}\xi_{2}\xi_{1}\xi_{0}+\psi_{3}\xi_{3}\xi_{1}\xi_{0}-\psi_{1}\xi_{3}\xi_{2}\xi_{1}\\\nonumber
       &&-\psi_{3}\psi_{2}\psi_{0}\xi_{0}+\psi_{3}\psi_{1}\psi_{0}\xi_{3}+\psi_{2}\psi_{1}\psi_{0}\xi_{2}+\psi_{3}\psi_{2}\psi_{1}\xi_{1})
\end{eqnarray}
and
\begin{eqnarray}
{\cal L}_3^{(1)} &=&\Phi\,g+A_x\dot{x}+A_y\dot{y}+\\\nonumber
&&\Phi_{x}(\psi_{0}\xi_{1}+\psi_{1}\xi_{0})+\Phi_{y}(\psi_{1}\xi_{1}-\psi_{0}\xi_{0})
-\Phi_{z}(\psi_{1}\psi_{0}+\xi_{1}\xi_{0})\ .
\end{eqnarray}

\newpage

\section{Appendix: Action for the (5,8,3) supermultiplet}

\noindent
The complete Lagrangian for the (5,8,3) multiplet reads
{\small
\begin{eqnarray}\nonumber
\Lcal_5 &=& \widetilde\Phi (\dot{v}_{1}^{2}+\dot{v}_{2}^{2}+\dot{w}^{2}+\dot{w}_{1}^{2}+\dot{w}_{2}^{2}+h_{1}^{2}+h_{2}^{2}+h_{3}^{2})+ \\ \nonumber
   && \widetilde\Phi (\dot{\lambda}_{0}\lambda_{0}+\dot{\lambda}_{1}\lambda_{1}+\dot{\lambda}_{2}\lambda_{2}+\dot{\lambda}_{3}\lambda_{3}+
   \dot{\chi}_{0}\chi_{0}+\dot{\chi}_{1}\chi_{1}+\dot{\chi}_{2}\chi_{2}+\dot{\chi}_{3}\chi_{3})+ \\ \nonumber
   && \widetilde\Phi_{v_{1}}[\dot{v}_{2}(\lambda_{0}\lambda_{1}+\lambda_{2}\lambda_{3}+
   \chi_{1}\chi_{0}+\chi_{2}\chi_{3})+h_{1}(\lambda_{1}\lambda_{3}+\lambda_{2}\lambda_{0}+
   \chi_{3}\chi_{1}+\chi_{2}\chi_{0}) \\ \nonumber
   && +h_{2}(\lambda_{2}\lambda_{1}+\lambda_{3}\lambda_{0}+
   \chi_{0}\chi_{3}+\chi_{2}\chi_{1})+h_{3}(\lambda_{0}\chi_{2}+\lambda_{3}\chi_{1}+\lambda_{2}\chi_{0}+\chi_{3}\lambda_{1})]+ \\ \nonumber
   && \widetilde\Phi_{v_{2}}[\dot{v}_{1}(\lambda_{1}\lambda_{0}+\lambda_{3}\lambda_{2}+\chi_{0}\chi_{1}+\chi_{3}\chi_{2})+
   h_{1}(\lambda_{2}\lambda_{1}+\lambda_{3}\lambda_{0}+\chi_{1}\chi_{2}+\chi_{3}\chi_{0}) \\ \nonumber
   &&+h_{2}(\lambda_{0}\lambda_{2}+\lambda_{3}\lambda_{1}+\chi_{3}\chi_{1}+\chi_{2}\chi_{0})+
     h_{3}(\chi_{1}\lambda_{2}+\lambda_{3}\chi_{0}+\lambda_{1}\chi_{2}+\lambda_{0}\chi_{3})]+ \\ \nonumber
   && \widetilde\Phi_{w}[\dot{w}_{1}(\lambda_{3}\lambda_{0}+\lambda_{1}\lambda_{2}+\chi_{0}\chi_{3}+\chi_{1}\chi_{2})+
      \dot{w}_{2}(\lambda_{0}\lambda_{2}+\lambda_{3}\lambda_{1}+\chi_{2}\chi_{0}+\chi_{1}\chi_{3}) \\ \nonumber
      &&+h_{1}(\chi_{2}\lambda_{3}+\chi_{1}\lambda_{0}+\chi_{0}\lambda_{1}+\lambda_{2}\chi_{3})+
      h_{2}(\chi_{1}\lambda_{1}+\lambda_{2}\chi_{2}+\lambda_{0}\chi_{0}+\lambda_{3}\chi_{3}) \\ \nonumber
      && +h_{3}(\lambda_{1}\lambda_{0}+\lambda_{2}\lambda_{3}+\chi_{1}\chi_{0}+\chi_{3}\chi_{2})]+ \\ \nonumber
   && \widetilde\Phi_{w_{1}}[\dot{w}(\lambda_{0}\lambda_{3}+\lambda_{2}\lambda_{1}+\chi_{2}\chi_{1}+\chi_{3}\chi_{0})+
      \dot{w}_{2}(\lambda_{2}\lambda_{3}+\lambda_{1}\lambda_{0}+\chi_{0}\chi_{1}+\chi_{2}\chi_{3}) \\ \nonumber
      && +h_{1}(\lambda_{3}\chi_{1}+\chi_{3}\lambda_{1}+\chi_{2}\lambda_{0}+\chi_{0}\lambda_{2})+
      h_{2}(\chi_{1}\lambda_{2}+\chi_{2}\lambda_{1}+\lambda_{0}\chi_{3}+\chi_{0}\lambda_{3}) \\ \nonumber
      && +h_{3}(\lambda_{2}\lambda_{0}+\lambda_{3}\lambda_{1}+\chi_{1}\chi_{3}+\chi_{2}\chi_{0})]+ \\ \nonumber
   && \widetilde\Phi_{w_{2}}[\dot{w}(\lambda_{3}\lambda_{1}+\lambda_{2}\lambda_{0}+\chi_{0}\chi_{2}+\chi_{3}\chi_{1})+
      \dot{w}_{1}(\lambda_{0}\lambda_{1}+\lambda_{3}\lambda_{2}+\chi_{1}\chi_{0}+\chi_{3}\chi_{2}) \\ \nonumber
      &&+h_{1}(\chi_{0}\lambda_{3}+\lambda_{1}\chi_{2}+\chi_{3}\lambda_{0}+\chi_{1}\lambda_{2})+
      h_{2}(\chi_{3}\lambda_{1}+\lambda_{2}\chi_{0}+\chi_{2}\lambda_{0}+\chi_{1}\lambda_{3}) \\\nonumber
      && +h_{3}(\lambda_{3}\lambda_{0}+\lambda_{1}\lambda_{2}+\chi_{3}\chi_{0}+\chi_{2}\chi_{1}) ]+ \\ \nonumber
&& \widetilde\Phi_{ww}(\lambda_{0}\chi_{0}\lambda_{1}\chi_{1}+\lambda_{2}\chi_{2}\lambda_{3}\chi_{3}+\chi_{0}\chi_{1}\chi_{2}\chi_{3})+ \\ \nonumber
   &&\widetilde\Phi_{v_{1}v_{1}}(\lambda_{0}\chi_{0}\chi_{2}\lambda_{2}+\lambda_{1}\chi_{1}\lambda_{3}\chi_{3}+
   \lambda_{0}\chi_{1}\chi_{2}\lambda_{3}-\chi_{0}\lambda_{1}\lambda_{2}\chi_{3}+\lambda_{0}\lambda_{1}\lambda_{2}\lambda_{3})+ \\ \nonumber
   && \widetilde\Phi_{v_{2}v_{2}}(\lambda_{0}\chi_{0}\chi_{3}\lambda_{3}+\lambda_{1}\chi_{1}\lambda_{2}\chi_{2}+
   \lambda_{0}\chi_{1}\lambda_{2}\chi_{3}-\chi_{0}\lambda_{1}\chi_{2}\lambda_{3}+\lambda_{0}\lambda_{1}\lambda_{2}\lambda_{3})+ \\ \nonumber
   && \widetilde\Phi_{w_{1}w_{1}}(\lambda_{0}\chi_{0}\chi_{2}\lambda_{2}+\lambda_{1}\chi_{1}\lambda_{3}\chi_{3}
   -\lambda_{0}\lambda_{1}\chi_{2}\chi_{3}+\lambda_{0}\chi_{1}\lambda_{2}\chi_{3}+\chi_{0}\chi_{1}\lambda_{2}\lambda_{3}+ \\\nonumber
   && -\chi_{0}\lambda_{1}\chi_{2}\lambda_{3}+\chi_{0}\chi_{1}\chi_{2}\chi_{3})+ \\ \nonumber
   && \widetilde\Phi_{w_{2}w_{2}}(\lambda_{0}\chi_{0}\chi_{3}\lambda_{3}+\lambda_{1}\chi_{1}\lambda_{2}\chi_{2}
   -\lambda_{0}\lambda_{1}\chi_{2}\chi_{3}+\lambda_{0}\chi_{1}\chi_{2}\lambda_{3}+\chi_{0}\chi_{1}\lambda_{2}\lambda_{3}+ \\\nonumber
   && -\chi_{0}\lambda_{1}\lambda_{2}\chi_{3}+\chi_{0}\chi_{1}\chi_{2}\chi_{3})+ \\ \nonumber
   && \widetilde\Phi_{v_{1}v_{2}}(\lambda_{0}\chi_{0}\chi_{2}\lambda_{3}-\lambda_{0}\chi_{0}\lambda_{2}\chi_{3}-
   \lambda_{0}\chi_{1}\chi_{2}\lambda_{2}+\lambda_{0}\chi_{1}\chi_{3}\lambda_{3}+\chi_{0}\lambda_{1}\lambda_{2}\chi_{2}\\\nonumber
   &&+\chi_{0}\lambda_{1}\chi_{3}\lambda_{3}+\lambda_{1}\chi_{1}\chi_{2}\lambda_{3}-\lambda_{1}\chi_{1}\lambda_{2}\chi_{3})+ \\ \nonumber
   && \widetilde\Phi_{v_{1}w}(-\lambda_{0}\chi_{0}\lambda_{1}\lambda_{2}-\lambda_{0}\chi_{0}\chi_{1}\chi_{2}+
      \lambda_{0}\lambda_{3}\lambda_{1}\chi_{1}+\lambda_{0}\lambda_{3}\chi_{2}\lambda_{2}+\chi_{0}\chi_{3}\chi_{1}\lambda_{1}\\\nonumber
    &&+\chi_{0}\chi_{3}\lambda_{2}\chi_{2}+\lambda_{1}\lambda_{2}\chi_{3}\lambda_{3}+\chi_{1}\chi_{2}\chi_{3}\lambda_{3}) +\\ \nonumber
   && \widetilde\Phi_{v_{1}w_{1}}(\lambda_{0}\lambda_{1}\chi_{2}\lambda_{3}-\lambda_{0}\chi_{1}\lambda_{2}\lambda_{3}-\chi_{0}\lambda_{1}\lambda_{2}\lambda_{3}-
   \lambda_{0}\lambda_{1}\lambda_{2}\chi_{3}-\chi_{0}\lambda_{1}\chi_{2}\chi_{3}\\\nonumber
   &&+\chi_{0}\chi_{1}\lambda_{2}\chi_{3}-\chi_{0}\chi_{1}\chi_{2}\lambda_{3}-\lambda_{0}\chi_{1}\chi_{2}\chi_{3})+ \\ \nonumber
   && \widetilde\Phi_{v_{1}w_{2}}(\lambda_{0}\chi_{0}\chi_{2}\chi_{3}+\lambda_{0}\chi_{0}\lambda_{2}\lambda_{3}+
   \chi_{0}\chi_{1}\chi_{2}\lambda_{2}+\chi_{0}\chi_{1}\lambda_{3}\chi_{3}+\lambda_{0}\lambda_{1}\lambda_{2}\chi_{2}+\\\nonumber
   &&\lambda_{0}\lambda_{1}\chi_{3}\lambda_{3}+\lambda_{1}\chi_{1}\lambda_{2}\lambda_{3}+\lambda_{1}\chi_{1}\chi_{2}\chi_{3})+ \\ \nonumber
   && \widetilde\Phi_{v_{2}w}(\lambda_{0}\chi_{0}\chi_{3}\chi_{1}+\lambda_{0}\chi_{0}\lambda_{3}\lambda_{1}+\chi_{0}\chi_{2}\lambda_{1}\chi_{1}+
   \chi_{0}\chi_{2}\chi_{3}\lambda_{3}+\lambda_{0}\lambda_{2}\chi_{1}\lambda_{1}\\\nonumber
   &&+\lambda_{0}\lambda_{2}\lambda_{3}\chi_{3}+\lambda_{2}\chi_{2}\lambda_{3}\lambda_{1}+\chi_{2}\lambda_{2}\chi_{1}\chi_{3})+ \\ \nonumber
   && \widetilde\Phi_{v_{2}w_{1}}(-\lambda_{0}\chi_{0}\chi_{2}\chi_{3}-\lambda_{0}\chi_{0}\lambda_{2}\lambda_{3}+
   \chi_{0}\chi_{1}\chi_{2}\lambda_{2}+\chi_{0}\chi_{1}\lambda_{3}\chi_{3}+\lambda_{0}\lambda_{1}\lambda_{2}\chi_{2}\\\nonumber
   &&+\lambda_{0}\lambda_{1}\chi_{3}\lambda_{3}+\lambda_{1}\chi_{1}\lambda_{3}\lambda_{2}+\lambda_{1}\chi_{1}\chi_{3}\chi_{2})+ \\ \nonumber
   && \widetilde\Phi_{v_{2}w_{2}}(-\chi_{0}\lambda_{1}\lambda_{2}\lambda_{3}+\lambda_{0}\lambda_{1}\lambda_{2}\chi_{3}-
   \lambda_{0}\lambda_{1}\chi_{2}\lambda_{3}-\lambda_{0}\chi_{1}\lambda_{2}\lambda_{3}-\chi_{0}\chi_{1}\lambda_{2}\chi_{3}\\\nonumber
   &&-\lambda_{0}\chi_{1}\chi_{2}\chi_{3}+\chi_{0}\chi_{1}\chi_{2}\lambda_{3}-\chi_{0}\lambda_{1}\chi_{2}\chi_{3}) +\\ \nonumber
   && \widetilde\Phi_{w w_{1}}(-\lambda_{0}\chi_{0}\lambda_{1}\chi_{2}+\lambda_{0}\chi_{0}\chi_{1}\lambda_{2}-\chi_{0}\lambda_{3}\lambda_{1}\chi_{1}+
   \chi_{0}\lambda_{3}\lambda_{2}\chi_{2}-\lambda_{0}\chi_{3}\lambda_{1}\chi_{1}\\\nonumber
   &&+\lambda_{0}\chi_{3}\lambda_{2}\chi_{2}-\lambda_{3}\chi_{3}\lambda_{1}\chi_{2}+\lambda_{3}\chi_{3}\chi_{1}\lambda_{2})+ \\ \nonumber
   && \widetilde\Phi_{w w_{2}}(\lambda_{0}\chi_{0}\chi_{3}\lambda_{1}-\lambda_{0}\chi_{0}\lambda_{3}\chi_{1}
   -\chi_{0}\lambda_{2}\chi_{1}\lambda_{1}+\chi_{0}\lambda_{2}\chi_{3}\lambda_{3}+\lambda_{0}\chi_{2}\lambda_{1}\chi_{1}\\\nonumber
   &&+\lambda_{0}\chi_{2}\chi_{3}\lambda_{3}+\lambda_{1}\chi_{3}\chi_{2}\lambda_{2}-\chi_{1}\lambda_{3}\chi_{2}\lambda_{2})+ \\ \nonumber
   && \widetilde\Phi_{w_{1}w_{2}}(\lambda_{0}\chi_{0}\chi_{2}\lambda_{3}-\lambda_{0}\chi_{0}\lambda_{2}\chi_{3}+
   \chi_{0}\lambda_{1}\chi_{2}\lambda_{2}+\chi_{0}\lambda_{1}\lambda_{3}\chi_{3}+\lambda_{0}\chi_{1}\chi_{2}\lambda_{2}\\
   &&+\lambda_{0}\chi_{1}\lambda_{3}\chi_{3}+\chi_{2}\lambda_{3}\lambda_{1}\chi_{1}+\lambda_{2}\chi_{3}\chi_{1}\lambda_{1}) \ .
\end{eqnarray}
}

\bigskip

\section{Appendix: ${\cal N}{=}4$ duality}

\noindent
It is instructive to display the simpler case of ${\cal N}{=}4$ duality. 
Since only the (1,4,3) multiplet allows for an inhomogeneous deformation,
we concentrate on the $d{=}1$ / $d{=}3$ duality and the coupling of
these two multiplets.

Like in the ${\cal N}{=}8$ cases, the ${\cal N}{=}4$ Lagrangians have the form
\begin{equation}
\Lcal_d \= \Phi\,\delta_{ab}\dot{x}^a\dot{x}^b\ +\ \ldots\ .
\end{equation}
Scale ($D$) and special conformal ($K$) invariance require
\begin{equation}
\Phi\=r^\beta\,Y({\rm angles}) \und \dot{\Phi}\,r^2 = \sfrac{\rm d}{{\rm d} t}Z
\quad\for r^2=x^a x^a\ ,
\end{equation}
with some exponent~$\beta$ and functions $Y$ and~$Z$.
It follows that $Z=\sfrac{c}{c+2}r^{\beta+2}Y$ and $Y={\rm constant}$.

Let us denote the components of the two multiplets by
\begin{equation}
\begin{cases}
(1,4,3): & x;\psi_0,\psi_1,\psi_2,\psi_3;f_1,f_2,f_3 \\[8pt]
(3,4,1): & v_1,v_2,v_3;\la_0,\la_1,\la_2,\la_3;h
\end{cases}
\end{equation}
and assign scaling dimensions ($i=0,1,2,3$ and $a=1,2,3$)
\begin{equation}
[x,\psi_i,f_a]=-1,-\sfrac12,0 \und [v_a,\la_i,h]=1,\sfrac32,2\ ,
\end{equation}
so that the conformal factors for a dimensionless action become
\begin{equation}
\Phi\=x^{-1} \und \widetilde{\Phi}\=v^{-3} \with v^2=v_a v_a\ .
\end{equation}
The bosonic target space is therefore the product of a (half) line with a
three-dimensional cone.
The supersymmetry transformations are given by\\
\begin{minipage}{7cm}
\begin{eqnarray*} 
\begin{array}{|c|c|c|c|c|}\hline
  & Q_{1} & Q_{2} & Q_{3} & Q_{4} \\\hline
  & & & &  \\[-10pt]
  x & \psi_1 & \psi_2 & \psi_3 & \psi_0 \\\hline
  & & & &  \\[-10pt]
  \psi_0 & f_1 & f_2 & f_3 & \dot{x} \\
  \psi_1 & \dot{x} & f_3{+}c & -f_2 & -f_1 \\
  \psi_2 & -f_3{-}c & \dot{x} & f_1 & -f_2 \\
  \psi_3 & f_2 & -f_1 & \dot{x} & -f_3 \\\hline
   & \qquad\quad & \qquad\quad & \qquad\quad & \qquad\quad \\[-10pt]
  f_1 & \dot{\psi}_0 & -\dot{\psi}_3 & \dot{\psi}_2 & -\dot{\psi}_1 \\
  f_2 & \dot{\psi}_3 & \dot{\psi}_0 & -\dot{\psi}_1 & -\dot{\psi}_2 \\
  f_3 & -\dot{\psi}_2 & \dot{\psi}_1 & \dot{\psi}_0 & -\dot{\psi}_3 \\\hline
\end{array}
\end{eqnarray*}
\end{minipage}
\begin{minipage}{9cm}
\begin{eqnarray} 
\begin{array}{|c|c|c|c|c|}\hline
  & Q_{1} & Q_{2} & Q_{3} & Q_{4} \\\hline
  & & & &  \\[-10pt]
  v_1 & \la_0 & -\la_3 & \la_2 & -\la_1 \\
  v_2 & \la_3 & \la_0 & -\la_1 & -\la_2 \\
  v_3 & -\la_2 & \la_1 & \la_0 & -\la_3 \\\hline
  & & & &  \\[-10pt]
  \la_0 & \dot{v}_1 & \dot{v}_2 & \dot{v}_3 & h \\
  \la_1 & h & \dot{v}_3 & -\dot{v}_2 & -\dot{v}_1 \\
  \la_2 & -\dot{v}_3 & h & \dot{v}_1 & -\dot{v}_2 \\
  \la_3 & \dot{v}_2 & -\dot{v}_1 & h & -\dot{v}_3 \\\hline
   & \qquad\quad & \qquad\quad & \qquad\quad & \qquad\quad \\[-10pt]
  h & \dot{\la}_1 & \dot{\la}_2 & \dot{\la}_3 & \dot{\la}_0 \\\hline
\end{array}
\end{eqnarray}
\end{minipage}\\[12pt]
with inhomogeneous parameter~$c$.
The transformations can be written in terms of the quaternionic structure constants 
$\delta_{ab}$ and $\epsilon_{abc}$ (with $\epsilon_{123}=1$). 
We note that the two multiplets must have the same chirality to be coupled.
Therefore, the overall sign of $\epsilon_{123}$ in the second multiplet is fixed
in order to allow the supersymmetric pairing of the multiplets.
 
The superconformally invariant action of the coupled system 
is given as a sum of three terms,
\begin{equation} \label{N4tot}
\Lcal_{1+3}\=\Lcal_1\ +\ \Lcal_3\ +\ \g\Lcal_{1,3}\ ,
\end{equation}
with
\begin{equation}
\Lcal_1 \= Q_4 Q_3 Q_2 Q_1 F(x) \und 
\Lcal_3 \= Q_4 Q_3 Q_2 Q_1 \widetilde{F}(\vec v)\ .
\end{equation}
The supersymmetric pairing term reads ($a=1,2,3$ and $i=0,1,2,3$)
\begin{eqnarray}\label{N4res1}
{\cal L}_{1,3}&=& {\cal L}_{1,3}^{(0)} +{\cal L}_{1,3}^{(1)}+{\cal L}_{1,3}^{(2)}\ ,\nonumber\\
{\cal L}_{1,3}^{(0)} &=& x\,h -f_a v_a +\psi_i\lambda_i\ ,\nonumber\\
{\cal L}_{1,3}^{(1)} &=& \omega_1\lambda_1+\omega_2\lambda_2\ ,\nonumber\\
{\cal L}_{1,3}^{(2)} &=& e\,h\ ,
\end{eqnarray} 
where the extra constants $\omega_1$, $\omega_2$ and $e$ have been added,
with scaling dimensions $[\omega_1]=[\omega_2]=-\sfrac{1}{2}$ and $[e]=-1$. 
The supersymmetry transformations of the constant (1,2,1) multiplet are
\begin{eqnarray} 
\begin{array}{|c|c|c|c|c|}\hline
  & Q_{1} & Q_{2} & Q_{3} & Q_{4} \\\hline
  & & & &  \\[-10pt]
 e & \omega_1 & \omega_2 & 0 & 0 \\\hline
  & & & &  \\[-10pt]
  \omega_1 & 0 & -c & 0 & 0 \\
  \omega_2 & c & 0 & 0 & 0 \\\hline
   & \qquad\quad & \qquad\quad & \qquad\quad & \qquad\quad \\[-10pt]
  c & 0 & 0 & 0 & 0 \\\hline
\end{array}
\end{eqnarray}

An alternative coupling possibility is the following,
\begin{eqnarray}\label{N4res2}
{\cal L}_{1,3}&=& {\cal L}_{1,3}^{(0)} +
{\cal L}_{1,3}^{(0')}+{\cal L}_{1,3}^{(1')}+{\cal L}_{1,3}^{(2')}\ ,\nonumber\\
{\cal L}_{1,3}^{(0)} &=& x\,h -f_a v_a +\psi_i\lambda_i\ ,\nonumber\\
{\cal L}_{1,3}^{(0')} &=& -c\,v_3\ ,\nonumber\\
{\cal L}_{1,3}^{(1')} &=& \omega_0\lambda_0+\omega_3\lambda_3\ ,\nonumber\\
{\cal L}_{1,3}^{(2')} &=& e' h\ ,
\end{eqnarray} 
where the extra constants $\omega_0$, $\omega_3$ and $e'$ have been added,
with scaling dimensions $[\omega_0]=[\omega_3]=-\sfrac{1}{2}$ and $[e']=-1$. 
The supersymmetry transformations of this constant (1,2,1) multiplet are
\begin{eqnarray} 
\begin{array}{|c|c|c|c|c|}\hline
  & Q_{1} & Q_{2} & Q_{3} & Q_{4} \\\hline
  & & & &  \\[-10pt]
 e' & 0 & 0 & \omega_3 & \omega_0  \\\hline
  & & & &  \\[-10pt]
  \omega_0 & 0 & 0 & c & 0 \\
  \omega_3 & 0 & 0 & 0 & -c \\\hline
   & \qquad\quad & \qquad\quad & \qquad\quad & \qquad\quad \\[-10pt]
  c & 0 & 0 & 0 & 0 \\\hline
\end{array}
\end{eqnarray}

The Lagrangians of the one- and three-dimensional systems read
\begin{eqnarray}
{\cal L}_1 &=& \Phi\,\bigl\{
\dot{x}^2+f_a^2+\dot{\psi}_0\psi_0+\dot{\psi}_a\psi_a \bigr\} \nonumber\\
&+& \Phi_x \bigl\{
\psi_0\psi_af_a+\sfrac{1}{2}\epsilon_{abc}\psi_a\psi_b f_c\bigr\}\ +\
\Phi_{xx}\bigl\{\sfrac{1}{6}\epsilon_{abc}\psi_0\psi_a\psi_b\psi_c\bigr\} \nonumber\\
&+&  c\,\Phi\,f_3\ +\ c\,\Phi_x \psi_0\psi_3
\end{eqnarray}
and
\begin{eqnarray}
{\cal L}_3 &=& \widetilde{\Phi}\,\bigl\{
\dot{v}_a^2+h^2+\dot{\lambda}_0\lambda_0+\dot{\lambda}_a\lambda_a \bigr\} \nonumber\\
 &+&\widetilde{\Phi}_a \bigl\{
 \la_a\la_b\dot{v}_b +\epsilon_{abc}(\sfrac12\la_b\la_c h-\la_0\la_b\dot{v}_c) \bigr\} +
 \sfrac16\Delta\widetilde{\Phi}\,\epsilon_{abc}\lambda_0\lambda_a\lambda_b\lambda_c\ ,
\end{eqnarray}
where
\begin{equation}
\Phi\=F_{xx} \und \widetilde{\Phi}\=\Delta\widetilde{F}\ \equiv\ \widetilde{F}_{aa}\ ,
\end{equation}
respectively.

Finally we add Fayet-Iliopoulos terms which are superconformal (not just supersymmetric) invariants
and introduce dimensionful constants $\mu_a$ and $\nu$,
\begin{equation}
{\cal L}_{\text{FI}} \= \mu_a f_a - \nu\,h\ ,
\end{equation}
with $[\mu_a]=1$ and $[\nu ]=-1$.
The supersymmetry transformations act trivially on $\mu_a$ and~$\nu$.

Setting all fermions to zero, the total bosonic Lagrangian 
based on (\ref{N4tot}) with (\ref{N4res1}) becomes
\begin{equation}
\Lcal_{1+3}'\big|\=
\Phi\,(\dot{x}^2+f_a^2)
+\widetilde{\Phi}\,(\dot{v}_a^2+h^2)
- (\g v_a-\mu_a-c\delta_{a3}\Phi)\,f_a
+(\g x+\g e-\nu)\,h \ .
\end{equation}
If we use (\ref{N4res2}) instead, an additional term $-\g c\,v_3$ appears.
Eliminating the auxiliary fields via
\begin{equation}
f_a \= \sfrac12\Phi^{-1} (\g v_a-\mu_a-c\delta_{a3}\Phi) \und 
h \= -\sfrac12\widetilde{\Phi}^{-1} (\g x+\g e-\nu) ,
\end{equation}
we arrive at
\begin{equation}
\Lcal_{1+3}''\big|\=
\Phi\,\dot{x}^2+\widetilde{\Phi}\,\dot{v}_a^2-
\sfrac14\Phi^{-1}\bigl(\g v_a-\mu_a-c\delta_{a3}\Phi\bigr)^2-
\sfrac14\widetilde{\Phi}^{-1}\bigl(\g(x{+}e)-\nu\bigr)^2\ ,
\end{equation}
where the Lagrange multiplier~$e$ only ensures that the zero mode $\overline{h}$
vanishes. Hence, its value is 
$e=-\overline{(\widetilde{\Phi}^{-1}x-\nu/\g)}\big/\overline{\widetilde{\Phi}^{-1}}$.
Specializing to $\Phi=x^{-1}$ and $\widetilde{\Phi}=v^{-3}$, one gets
\begin{equation}\label{N4L2}
\Lcal_{1+3}''\big|\=
x^{-1}\dot{x}^2+v^{-3}\dot{v}_a^2-
\sfrac14 x\,\bigl(\g v_a-\mu_a-c\delta_{a3}x^{-1}\bigr)^2-
\sfrac14 v^3\bigl(\g(x{+}e)-\nu\bigr)^2\ .
\end{equation}
In order to interpret this Lagrangian, we pass to standard kinetic terms (up to a
factor of~$\sfrac12$) by changing the radial coordinates via
\begin{equation}
x = \sfrac14 \rho^2 \und v = 4\sigma^{-2} \quad\with [\rho]=[\sigma]=-\sfrac12
\end{equation}
and arrive at
\begin{equation}\label{N4L3}
\Lcal_{1+3}^{\rm cone}\big|\=
\dot{\rho}^2 + \dot{\sigma}^2 + 4\widetilde{\ell}^2\sigma^{-2}
-\sfrac1{16} \rho^2 \bigl(\g\sigma^{-2}\vec{e}_\sigma-\vec\mu-4c\rho^{-2}\vec{e}_3\bigr)^2
-\sigma^{-6} \bigl(\g(\rho^2{+}4e)-4\nu\bigr)^2\ ,
\end{equation}
where $\widetilde{\ell}$ is the angular momentum in $\sigma$ space, and
$\vec{e}_\sigma$ and $\vec{e}_3$ denote unit vectors in the $\sigma$ and $3$~directions, 
respectively. We find a rather complicated potential in the four-dimensional target.
If one employs the option~(\ref{N4res2}), then linear terms 
$-\g c\,v_3$ and $-4\g c\,\sigma^{-2}\vec{e}_3$
have to be added to (\ref{N4L2}) and~(\ref{N4L3}), respectively.

\bigskip

\noindent
{\bf Acknowledgments}

\noindent
This work was partially supported by CNPq and by the Deutsche Forschungsgemeinschaft 
under grant LE 838/12.
M.G.\ and O.L.\ acknowledge a PCI-BEV grant and hospitality at CBPF.

\end{document}